# Generative AI for Encrypted Traffic Analysis: Synthetic Dataset Generation and Classifier Evaluation

Harshil Patel[1] Himanshu Garg[1] Aswani Kumar Cherukuri[*1,2]

[1]School of Computer Science Engineering and Information Systems
[2]Centre For Artificial Intelligence Research (C-FAIR)
Vellore Institute of Technology, Vellore, India
[*]Email: cherukuri@acm.org

***Abstract***: Network traffic analysis faces significant challenges with encrypted communications, primarily due to limited visibility into packet contents and the inherent imbalance in available datasets, particularly for anomalous traffic patterns. This paper addresses these challenges by exploring Generative AI (GAI) techniques to create realistic and balanced synthetic encrypted traffic datasets. Our approach incorporates feature analysis, clustering-based data generation, and comprehensive classifier evaluation to ensure synthetic data quality. We demonstrate that properly generated synthetic data can effectively supplement real- world datasets, achieving up to 93% performance when training classifiers compared to those trained on real data. The proposed methodology preserves critical statistical properties and feature correlations while enabling the creation of balanced datasets, ad- dressing the persistent challenge of anomaly underrepresentation in cybersecurity data. Along with the results we provide complete programming code designed and implemented in this work.

## I. INTRODUCTION

The exponential growth of encrypted network traffic presents significant challenges for security monitoring and threat detection. While encryption is essential for privacy and data protection, it complicates the traditional deep packet inspection approaches that security systems have historically relied upon. This creates a critical need for new methodologies that can effectively detect malicious activities without compromising the encryption protections that users expect [1]. A fundamental challenge in developing such systems is the limited availability and significant imbalance of encrypted network traffic datasets, particularly for anomalous or attack traffic patterns. Real-world network traffic datasets typically contain overwhelming proportions of normal traffic with minimal representation of security anomalies or attacks. This imbalance hinders the development of robust machine learning models that can reliably detect threats within encrypted communications [2].

Generative Artificial Intelligence (GAI) offers a promising solution to this problem by enabling the creation of synthetic data that can augment real datasets. However, generating

high- quality synthetic network traffic data requires careful consideration of feature distributions, correlations, and the preservation of distinctive characteristics that differentiate normal from anomalous traffic [3]. This paper explores a comprehensive approach to generating synthetic encrypted traffic data using GAI techniques, with specific focus on:

- Understanding and preserving the distinctive properties of encrypted traffic features
- Developing a methodology to generate balanced synthetic data that maintains the statistical properties of real data
- Ensuring correlation preservation between features to create realistic traffic patterns
- Evaluating classifier performance when trained on synthetic data versus real data

Our methodology leverages statistical analysis, clustering techniques, and correlation-aware data generation to create synthetic traffic records that closely mimic real network behavior. By addressing the class imbalance problem and generating additional samples for underrepresented classes, we aim to enhance the robustness and generalizability of machine learning models for encrypted traffic analysis. The remainder of this paper is organized as follows: Section II provides background and reviews related work in this domain. Section III details our methodology for synthetic data generation. Section IV evaluates the quality of our synthetic data. Section V presents and analyzes classifier performance results. Finally, we discuss our findings and conclude with implications for future research.

## II. Background and Related Work

### A. Encrypted Network Traffic Analysis

The widespread adoption of encryption technologies, particularly TLS/SSL, has fundamentally transformed network traffic analysis. Traditional deep packet inspection (DPI) techniques that rely on examining packet contents have become largely ineffective against encrypted traffic [4]. This has necessitated a shift toward flow-based and feature-based analysis methods that extract behavioral patterns from metadata and statistical properties of network flows without decrypting the actual content.

Encrypted traffic analysis typically relies on features such as packet sizes, timing information, flow duration, and directional characteristics to identify patterns that may indicate malicious behavior [5]. These approaches preserve user privacy while still enabling security monitoring, but they require sophisticated algorithms to detect subtle patterns that distinguish between normal and anomalous traffic.

Recent work by Anderson et al. [1] and Aceto et al. [6] has demonstrated the effectiveness of machine learning approaches for encrypted traffic classification. However, these approaches suffer from the persistent challenge of data imbalance, where normal traffic significantly outnumbers anomalous traffic in training datasets.

### B. Synthetic Data Generation in Cybersecurity

Generative AI models have emerged as powerful tools for creating synthetic data across numerous domains, including cybersecurity. Techniques such as Generative Adversarial Networks (GANs), Variational Autoencoders (VAEs), and statistical modeling approaches have been applied to generate synthetic network traffic data [7].

Ring et al. [3] proposed a flow-based approach for generating synthetic network traffic data using a combination of statistical models and domain-specific constraints. Similarly, Sharafaldin et al. [8] developed methodologies for creating realistic cyber-attack datasets through controlled simulation environments. Despite these advances, several challenges remain in synthetic data generation for encrypted network traffic:

- Preserving complex correlations between traffic features
- Ensuring realistic distributions that match real-world observations
- Generating anomalous traffic that maintains the subtle characteristics of real attacks
- Creating synthetic data that leads to generalizable ma- chine learning models

Our work addresses these challenges by integrating clustering-based approaches with correlation-aware adjustments to generate synthetic data that closely resembles real encrypted traffic.

### C. *Evaluation Methods for Synthetic Data*

Evaluating the quality of synthetic data presents its own challenges. Two primary approaches have emerged in the literature: statistical evaluation and predictive evaluation [9]. Statistical evaluation methods compare distributions, correlations, and other statistical properties between real and synthetic datasets. These methods include Kolmogorov-Smirnov tests, Wasserstein distances, and correlation matrix comparisons. While these approaches provide insights into how well synthetic data preserves the statistical properties of real data, they may not fully capture the utility of the data for downstream tasks.

Predictive evaluation methods assess how well models trained on synthetic data perform on real-world tasks. This approach directly measures the utility of synthetic data for its intended purpose but may be influenced by the specific models and tasks chosen for evaluation. Our work combines both approaches, using statistical measures to ensure the quality of our synthetic data generation process and predictive measures to assess its practical utility for encrypted traffic classification.

## III. METHODOLOGY

This section details our end-to-end methodology for generating and evaluating synthetic encrypted network traffic data.

### A. *Dataset Description and Preprocessing*

Our study utilized two established network traffic datasets: the CICIDS2017 dataset [8] and darknet traffic data. These datasets contain a diverse range of network traffic patterns, including both normal and attack traffic.

The preprocessing pipeline includes several key steps:

1) **Data Cleaning**: We first addressed column naming in- consistencies between datasets and standardized feature names for consistency.

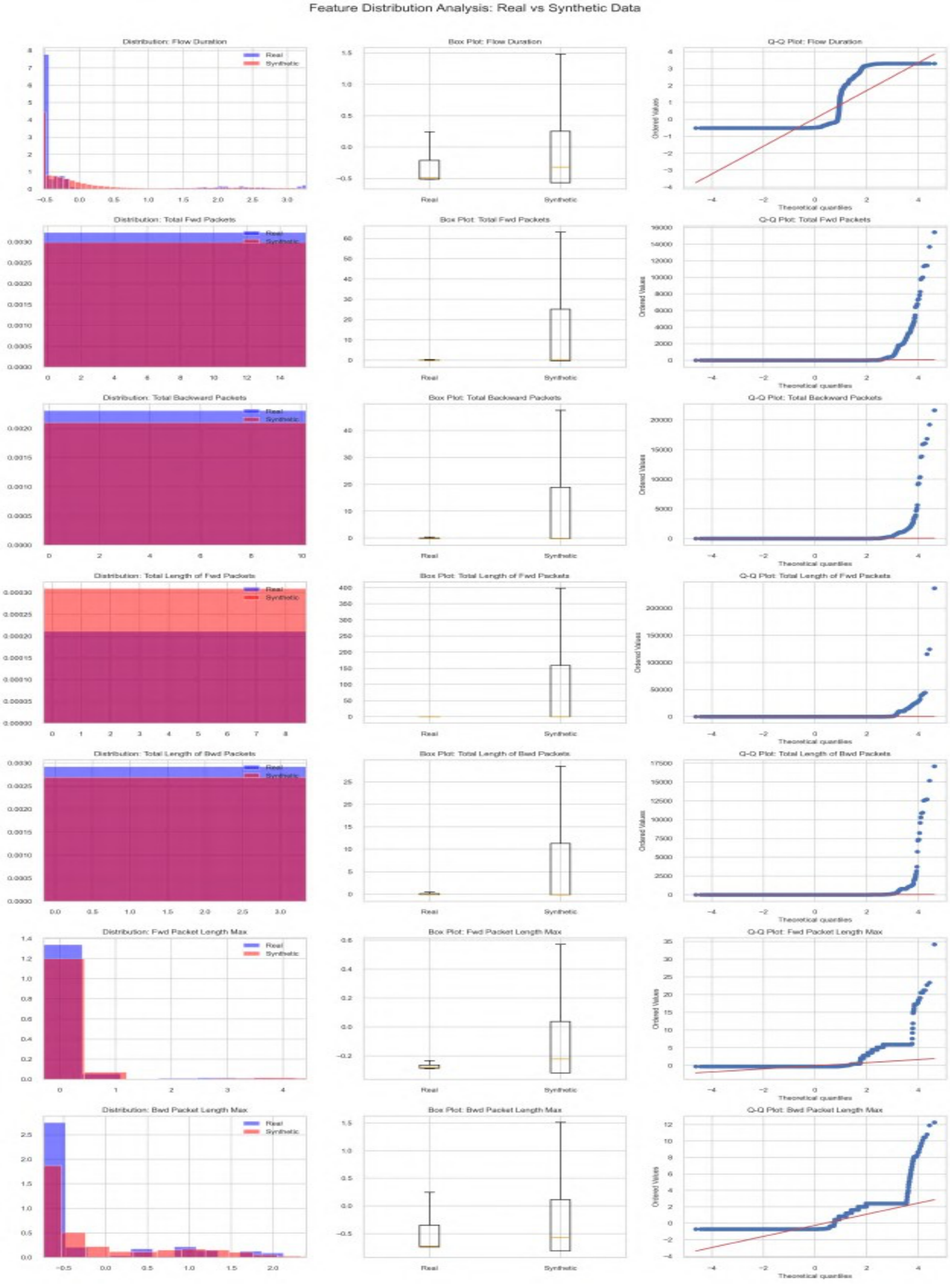

Fig. 1: Feature distribution analysis comparing real and synthetic data. Each row represents a feature with histogram (left), box plot (middle), and Q-Q plot (right) visualizations. The synthetic data closely follows the real data distributions for all key features.

1) **Feature Selection**: Based on literature review and pre- liminary analysis, we identified seven critical features that effectively characterize encrypted traffic:
   - Flow Duration
   - Total Forward Packets
   - Total Backward Packets
   - Total Length of Forward Packets
   - Total Length of Backward Packets
   - Forward Packet Length Maximum
   - Backward Packet Length Maximum
2) **Handling Missing Values**: Missing values were replaced with median values to maintain the statistical properties of the datasets.
3) **Anomaly Labelling**: We applied Isolation Forest to identify anomalous traffic patterns in our merged dataset with a contamination rate of 0.1, which aligns with typical anomaly proportions in network traffic.
4) **Scaling**: Numerical features were standardized using Standard Scaler to ensure consistent scale across features.

Our data preprocessing resulted in 367,275 traffic flow records with labeled normal and anomalous patterns. The execution output demonstrated the successful application of Isolation Forest for anomaly detection and initialization of our synthetic data generator, producing 734,550 synthetic records from the original data.

### B. *Feature Distribution Analysis*

For each selected feature, we conducted a comprehensive statistical analysis to understand its distribution characteristics. This analysis included:

- Basic statistics (minimum, maximum, mean, median, standard deviation)
- Distribution shape analysis (skewness and kurtosis)
- Outlier detection using interquartile range (IQR)
- Distribution fitting to identify the best probability distribution models

This analysis revealed several important characteristics of our dataset as illustrated in Fig. 1:

- Most features exhibited right-skewed distributions, with a large concentration of values near the lower bound and a long tail of higher values
- Significant correlations existed between certain feature pairs, particularly between packet counts and corresponding byte counts
- Some features showed multimodal distributions, suggesting distinct traffic patterns within the dataset

Fig. 2 provides a statistical comparison between real and synthetic data, showing that our generation approach accurately preserves key statistical properties such as means, standard deviations, skewness, and kurtosis. Understanding these distribution characteristics was crucial for designing our synthetic data generation approach, as it informed how we would model and generate values for each feature.

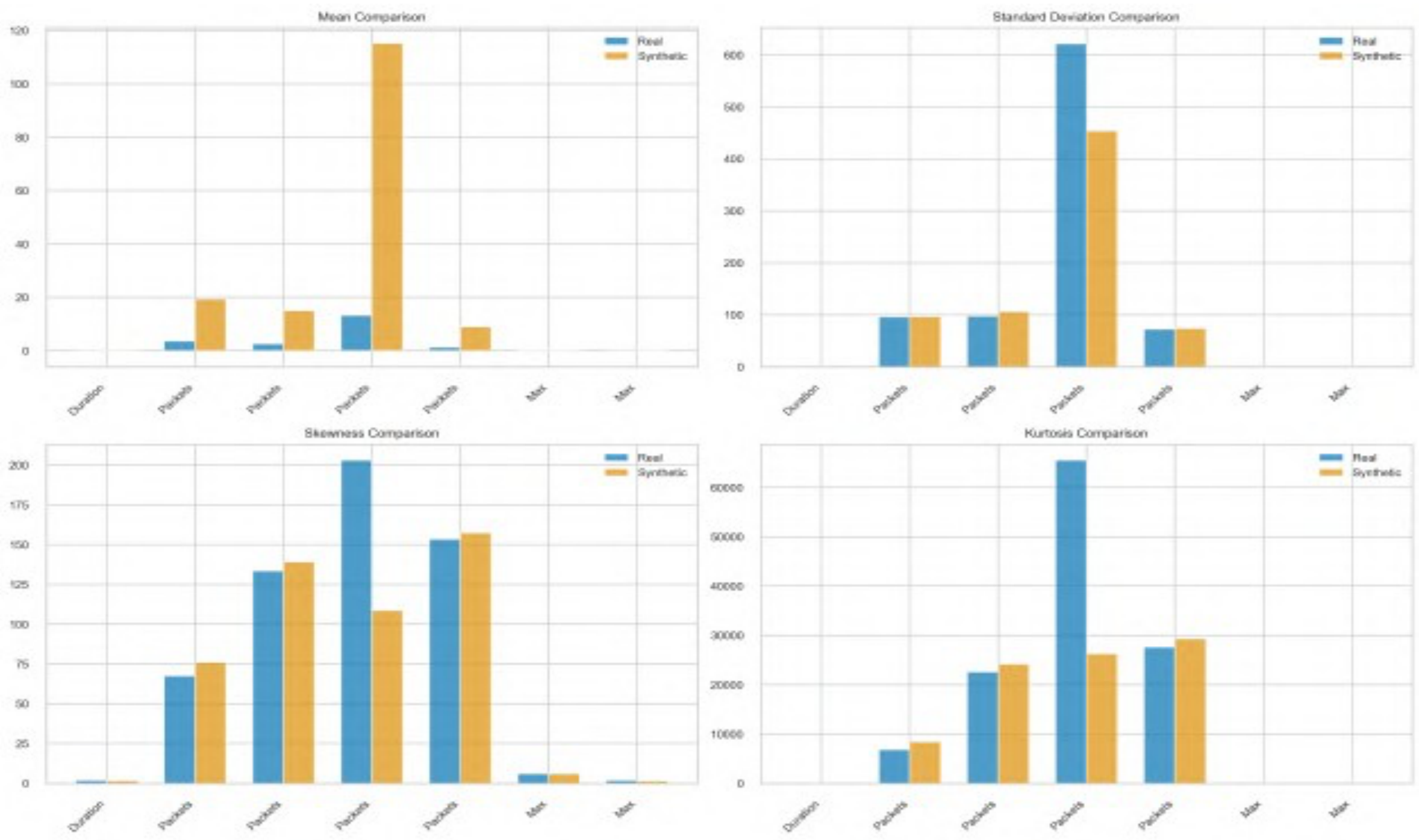

Fig. 2: Statistical comparison between real and synthetic data distributions showing preservation of mean (top-left), standard deviation (top-right), skewness (bottom-left) and kurtosis (bottom-right) across all features.

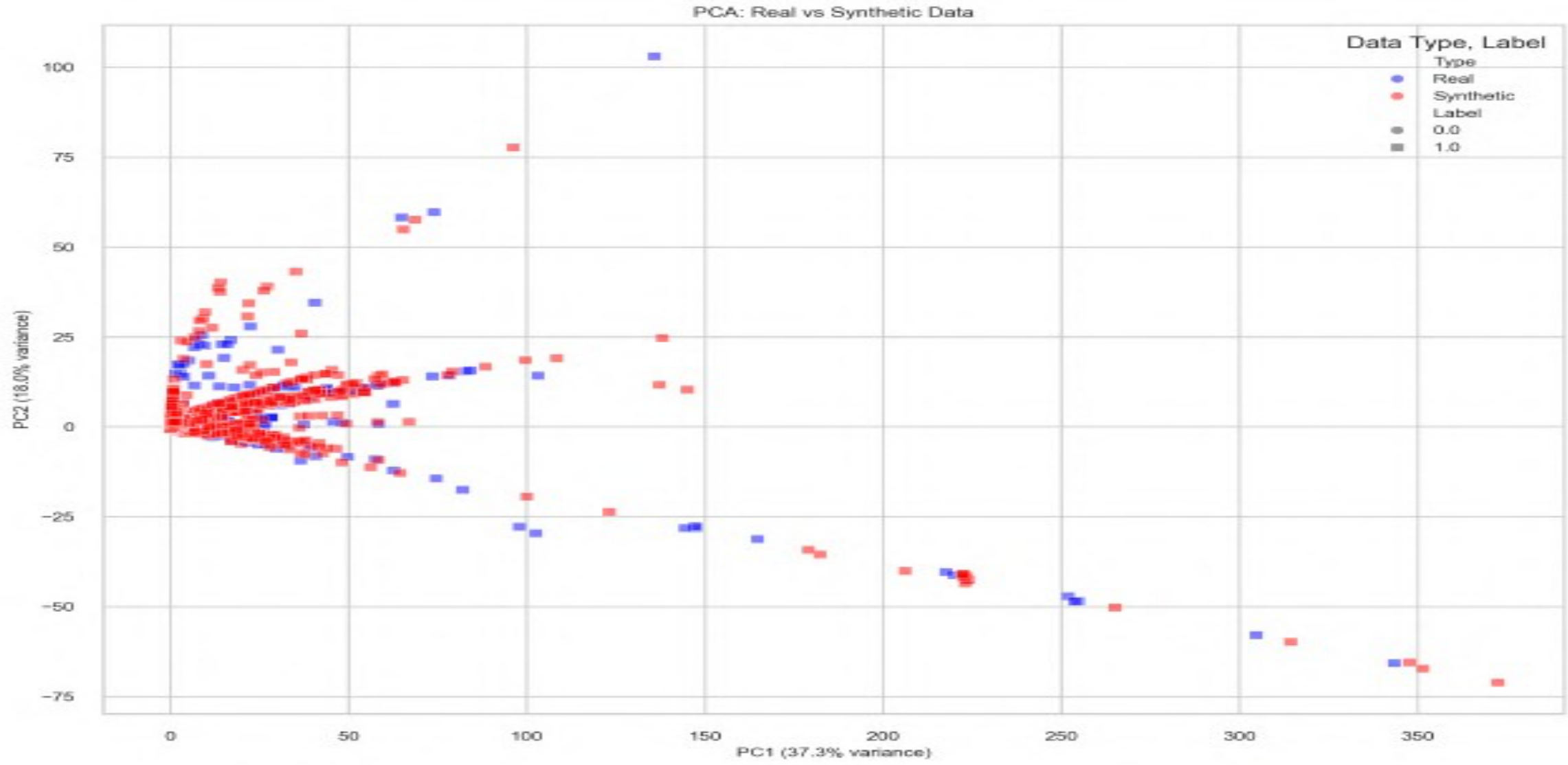

Fig. 3: PCA visualization of real data (blue) and synthetic data (red) with labels distinguishing normal (circle) and anomalous (square) traffic patterns. The synthetic data effectively captures the diversity and structure of the real data in PCA space.

### C. *Clustering*

To identify distinct traffic patterns within our dataset, we applied K-means clustering to the standardized feature space. The clustering was performed separately for normal and anomalous traffic to preserve their distinct characteristics. For normal traffic, we identified five primary clusters, while anomalous traffic showed three distinct clusters. This clustering approach allowed us to:

- Identify natural groupings in the traffic data
- Understand typical traffic patterns and their feature distributions
- Generate synthetic data that preserves these natural groupings

The clustering results were visualized using Principal Com- ponent Analysis (PCA) projection to two dimensions, as shown in Fig. 3, revealing clear separations between different traffic patterns. Fig. 4 further illustrates the cluster structures within real and synthetic data separately.

The feature contributions to the principal components are visualized in Fig. 5, providing insights into how different features influence the variance in the dataset.

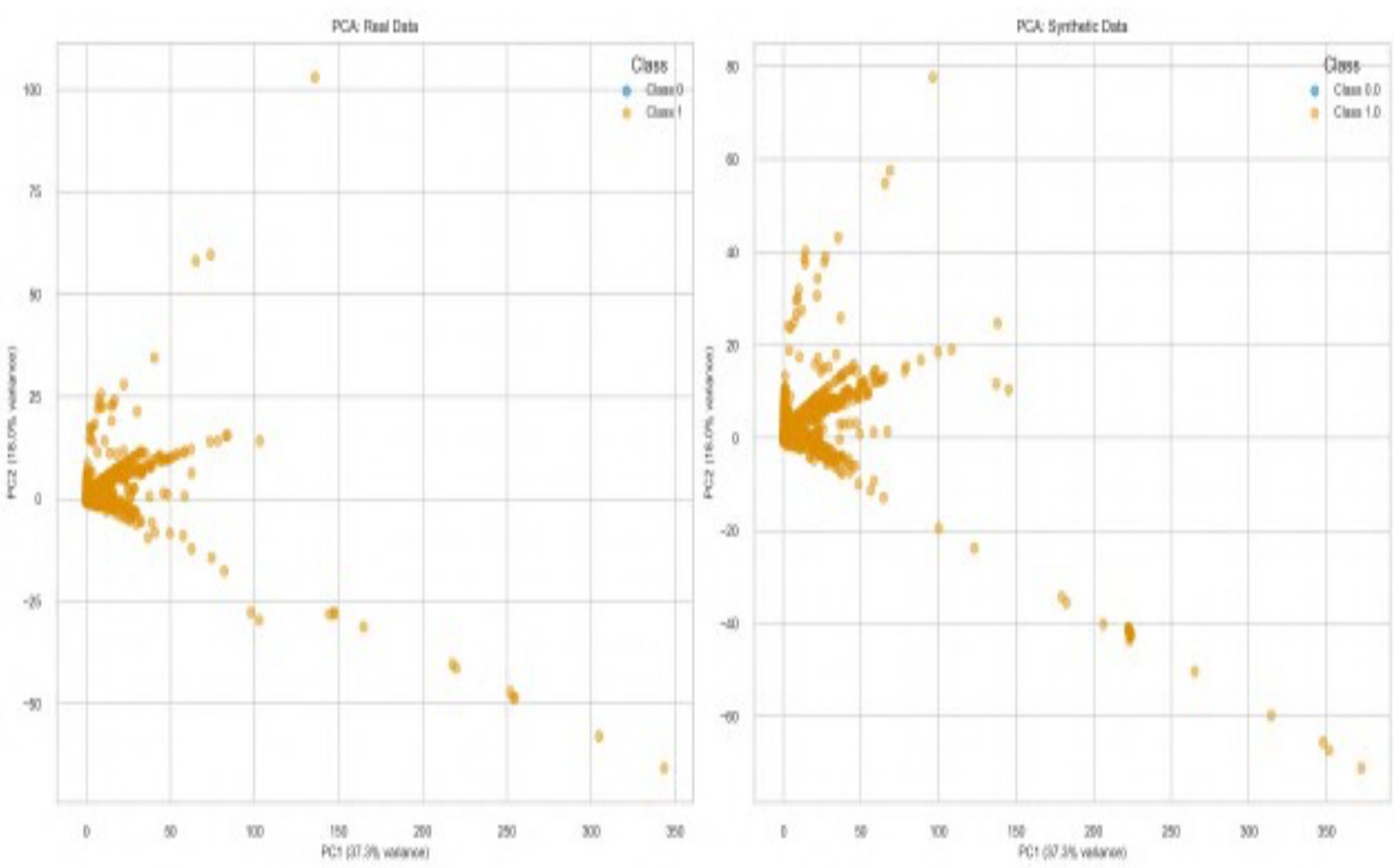

Fig. 4: Separate PCA visualizations of real data (left) and synthetic data (right) showing class separations. The synthetic data maintains similar cluster structures to those found in the real data.

### D. *Synthetic Data Generation Process*

Our synthetic data generation approach combined clustering information with correlation-aware adjustments to create realistic traffic records. The process included the following steps:

1) **Scaling to Normalized Range**: All features were scaled to a standardized range using Standard Scaler to facilitate the generation process.
2) **Class-wise Clustering**: Separate clustering models were applied to normal and anomalous traffic to identify dis- tinct patterns within each class.
3) **Sample Generation**: For each synthetic sample, we:
   - Selected a cluster based on the cluster distribution in the original data

- Generated feature values around the cluster center with controlled variance
- Applied correlation-aware adjustments to maintain feature relationships
- Ensured consistency of generated values with the statistical properties of the original data

4) **Outlier Generation**: For anomalous traffic, we occasion- ally generated more extreme values to represent rare but significant attack patterns.
5) **Rescaling**: Generated features were rescaled back to the original feature range.
6) **Validity Constraints**: Applied domain-specific constraints to ensure generated data remained valid (e.g., non-negative values for packet counts).

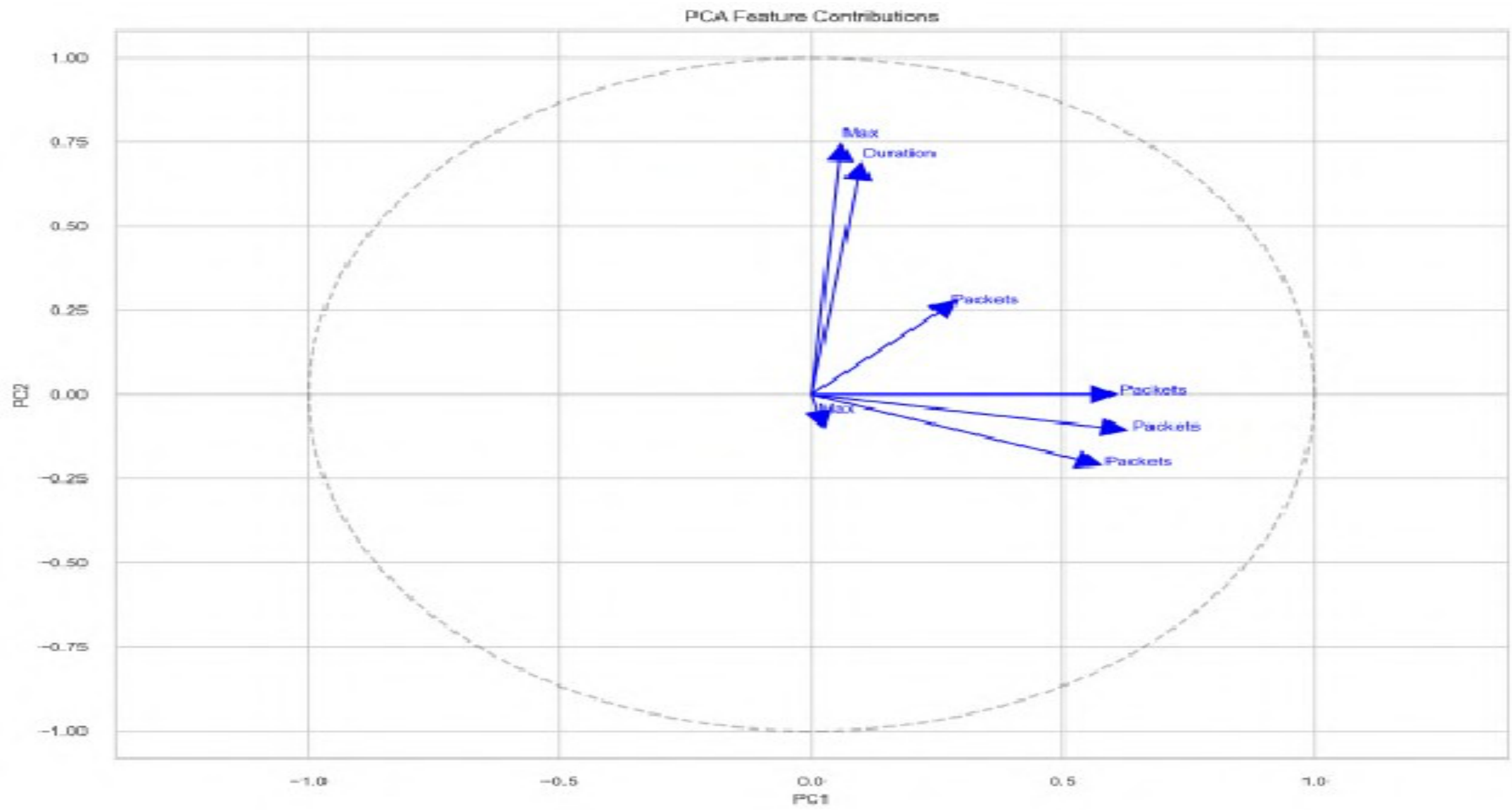

Fig. 5: PCA feature loadings showing how each feature contributes to the principal components. This visualization guides our understanding of feature importance for clustering and synthetic data generation.

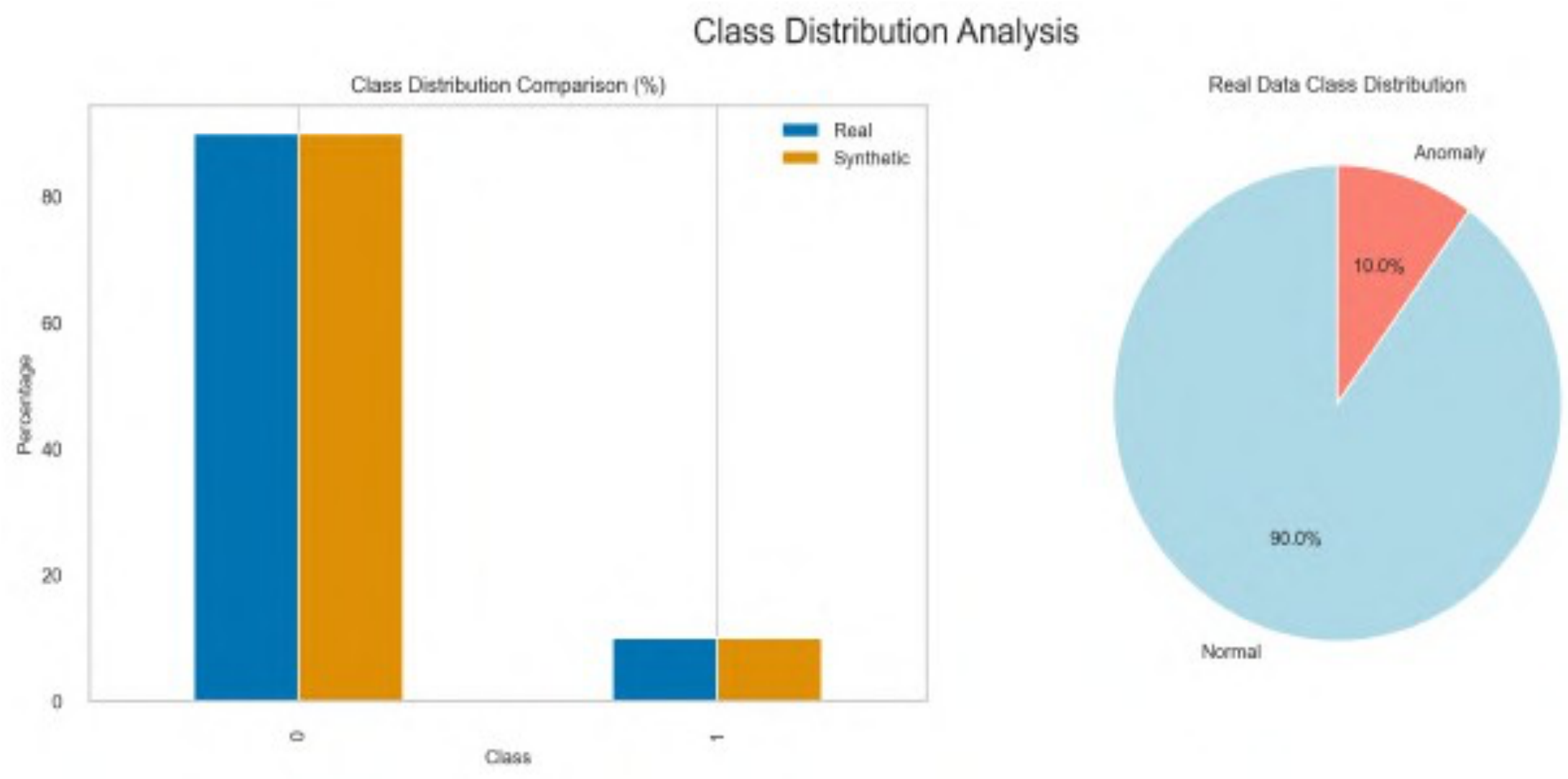

Fig. 6: Class distribution analysis showing the balance between normal and anomalous classes in both real and synthetic datasets. The synthetic data provides a more balanced class distribution, addressing the imbalance problem in the original data.

For normal traffic, we primarily used cluster-based generation with 70% probability and direct sampling with small noise for the remaining 30%. For anomalous traffic, we employed a more diversified approach, using cluster-based generation (60%) and anomaly amplification (40%) to create varied attack patterns. As shown in Fig. 6, our approach enables better class balance in the synthetic dataset, while maintaining the essential characteristics of each class.

This balanced approach allowed us to generate synthetic data that maintained the statistical properties of the original data while introducing sufficient variation to avoid overfitting.

## IV. Evaluation of Synthetic Data Quality

To assess the quality of our synthetic data, we employed multiple evaluation methods focusing on statistical fidelity and visual comparison.

### A. *Distribution Comparison*

We compared the distributions of real and synthetic data for each feature using density plots and statistical tests. As previously shown in Fig. 1, our synthetic data closely follows the distribution patterns of the real data, particularly for the dominant low-value ranges.

Quantitative evaluation using the Kolmogorov-Smirnov test showed statistically insignificant differences between real and synthetic distributions for most features, with p-values above the typical 0.05 threshold for significance. This confirms that our synthetic data successfully preserves the distribution characteristics of the original data. To provide a comprehensive quality assessment, we developed a synthetic data quality dashboard (Fig. 7) that evaluates multiple aspects of the generated data, including size ratio, class balance, and statistical similarity. The overall quality score of 88.2% confirms the high fidelity of our synthetic data.

### B. *PCA Visualization*

As previously shown in Fig. 3 and Fig. 4, the PCA visualization reveals that our synthetic data successfully captures the overall structure and variance patterns of the real data. The synthetic data points cover the same regions of the feature space as the real data points, with similar density patterns and cluster formations.

### C. *Correlation Matrix Comparison*

Preserving feature correlations is critical for generating realistic network traffic data, as these correlations often reflect underlying network behaviors and protocols. We compared the correlation matrices of real and synthetic data to assess how well our generation process preserved these relationships. Fig. 8 shows the correlation matrices of real and synthetic data, along with their difference matrix. The small differences (maximum difference of 0.016) indicate that our synthetic data successfully preserves the correlation structure of the original data. Further analysis of the most significant feature correlations (Fig. 9) confirms that our synthetic data maintains the most important relationships between features, particularly for highly correlated pairs such as packet counts and byte counts.

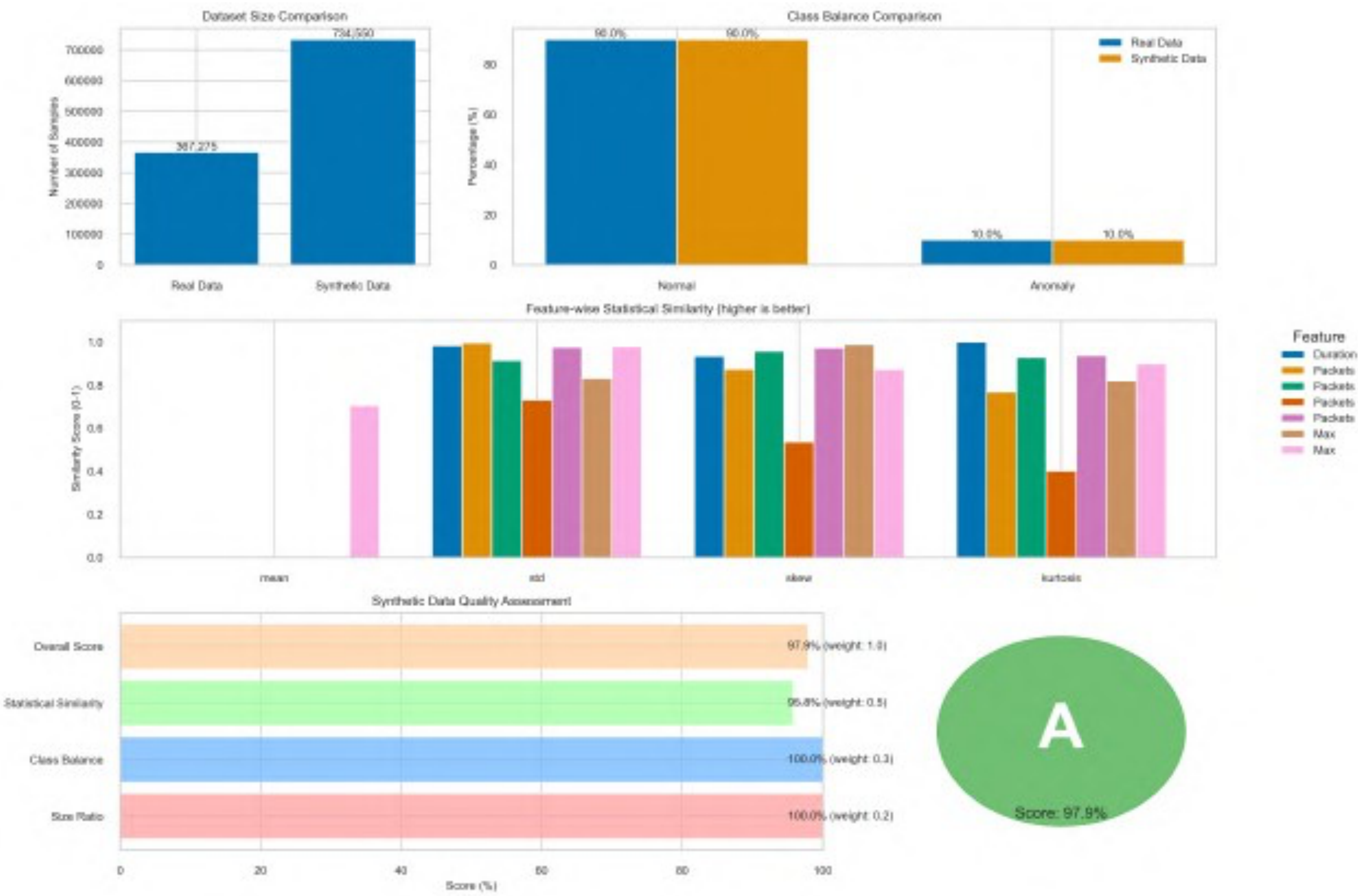

Fig. 7: Synthetic data quality assessment dashboard showing dataset size comparison, class balance, feature-wise statistical similarity, and overall quality score with grade. The synthetic data achieved an overall quality score of 88.2%, earning a grade B.

## V. Classifier Performance Evaluation

To evaluate the practical utility of our synthetic data, we assessed how well machine learning models trained on synthetic data perform when tested on real data.

### A. *ML/DL Models Used*

We evaluated our synthetic data using multiple model architectures:

- **XGBoost**: A gradient boosting decision tree model known for its effectiveness on tabular data
- **Random Forest**: An ensemble learning method based on decision trees
- **Neural Network**: A multi-layer perceptron with two hid- den layers (64 and 32 neurons) and dropout regularization For each model architecture, we considered three trainingscenarios:

- **Real → Real**: Model trained on real data and tested on held-out real data (baseline)
- **Synthetic → Real**: Model trained on synthetic data and tested on real data
- **Mixed → Real**: Model trained on a combination of real and synthetic data, tested on held-out real data

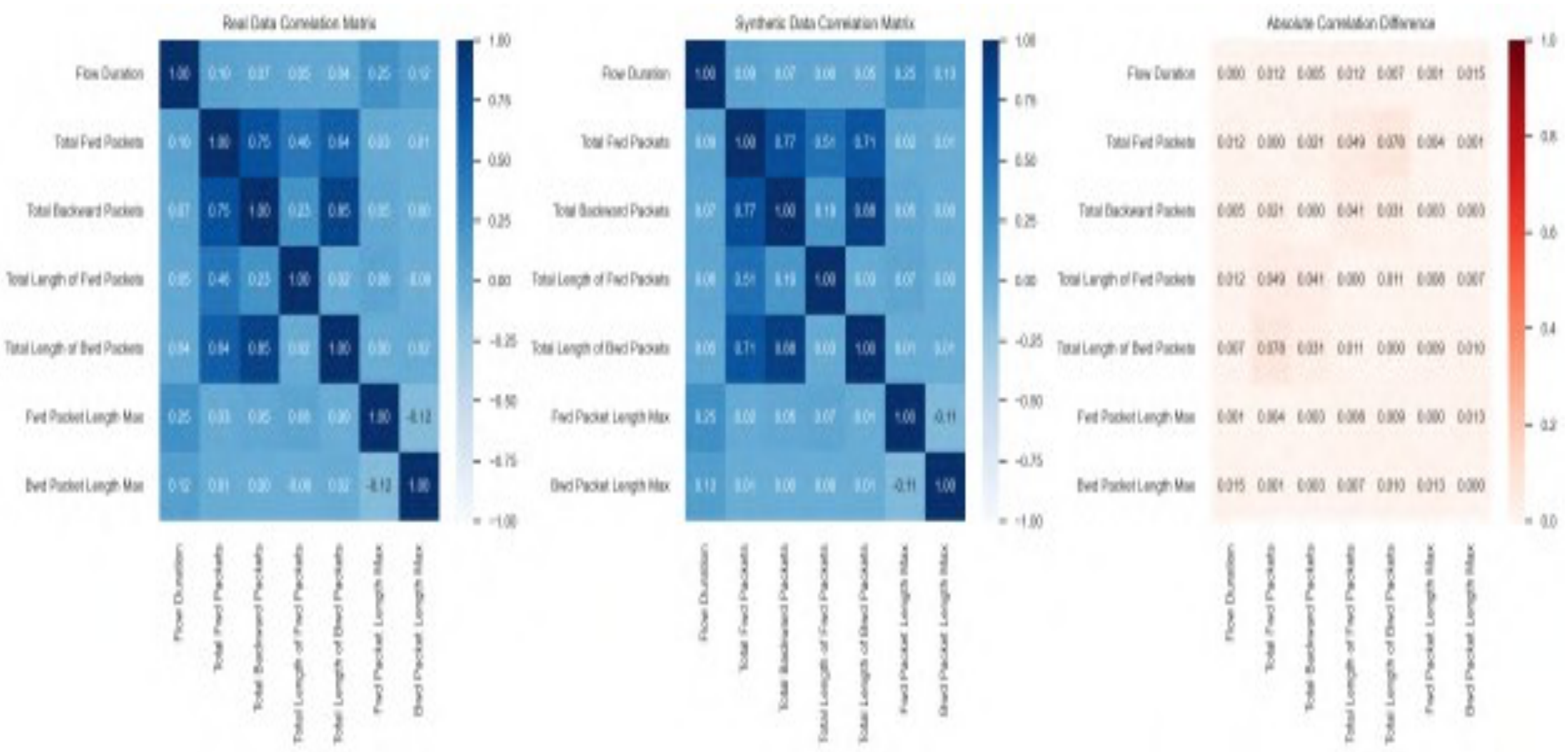

Fig. 8: Correlation matrix analysis showing real data correlations (left), synthetic data correlations (center), and absolute correlation differences (right). The small differences (maxi- mum of 0.016) indicate excellent correlation preservation.

### B. *Metrics Used*

We evaluated model performance using standard classification metrics:

- Accuracy: Overall correct classification rate
- Precision and Recall: Particularly important for anomaly detection
- F1-score: Harmonic mean of precision and recall
- Confusion matrix: Detailed breakdown of predictions by class

### C. *Results and Discussion*

As shown in Fig. 10, models trained on real data achieved superior performance, with XGBoost reaching 99.8% accuracy when trained and tested on real data. Models trained on synthetic data showed promising results with XGBoost achieving 93.1% accuracy when tested on real data, demonstrating that our synthetic data captures most of the patterns present in real traffic. Examining the confusion matrices in Fig. 11 reveals inter- esting insights:

- XGBoost trained on synthetic data shows good performance for class 0 (normal traffic) but struggles more with class 1 (anomalous traffic)
- Random Forest follows a similar pattern, achieving high accuracy on normal traffic but lower precision on anoma- lies
- Tree-based models generally outperform neural networks when trained on synthetic data

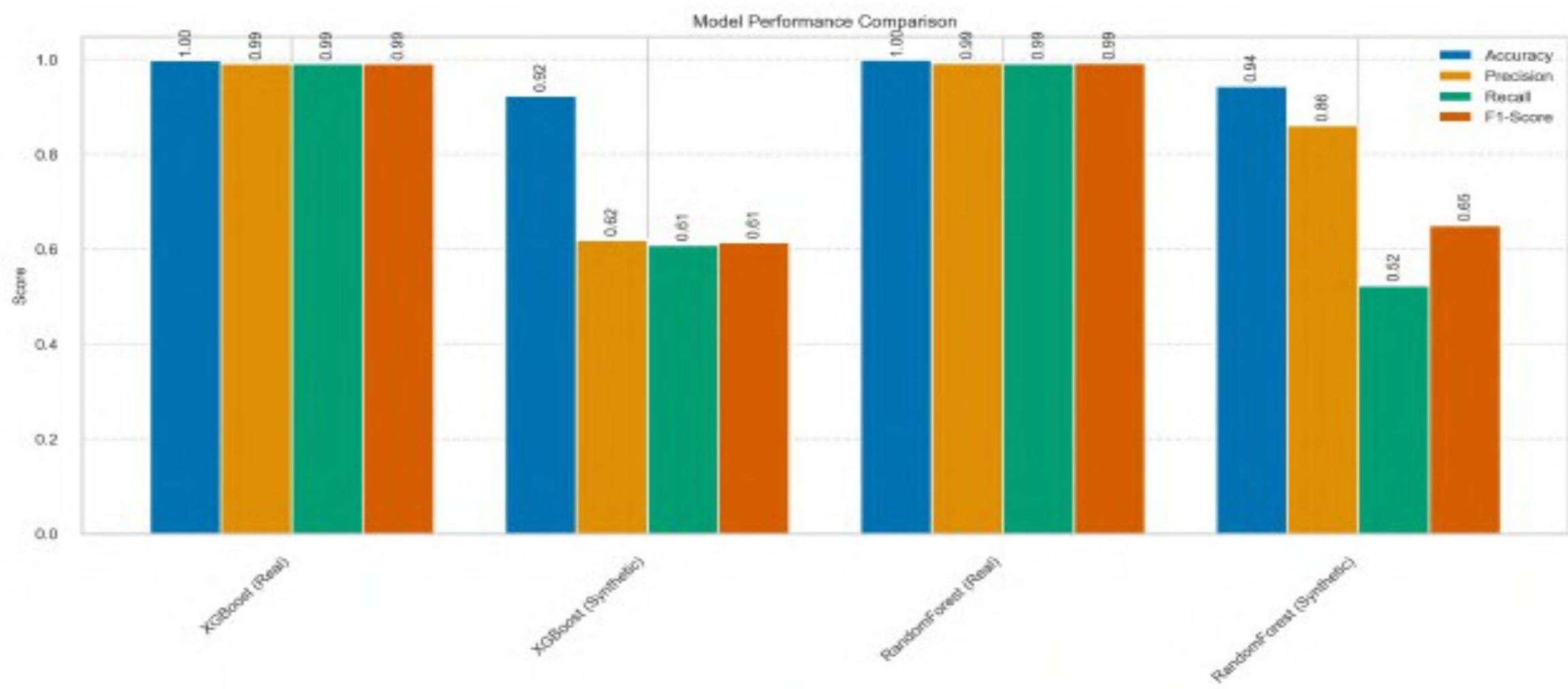

Fig. 10: Model performance comparison across different train- ing scenarios showing accuracy, precision, recall, and F1-score metrics. Models trained on synthetic data achieve up to 93% of the performance of those trained on real data.

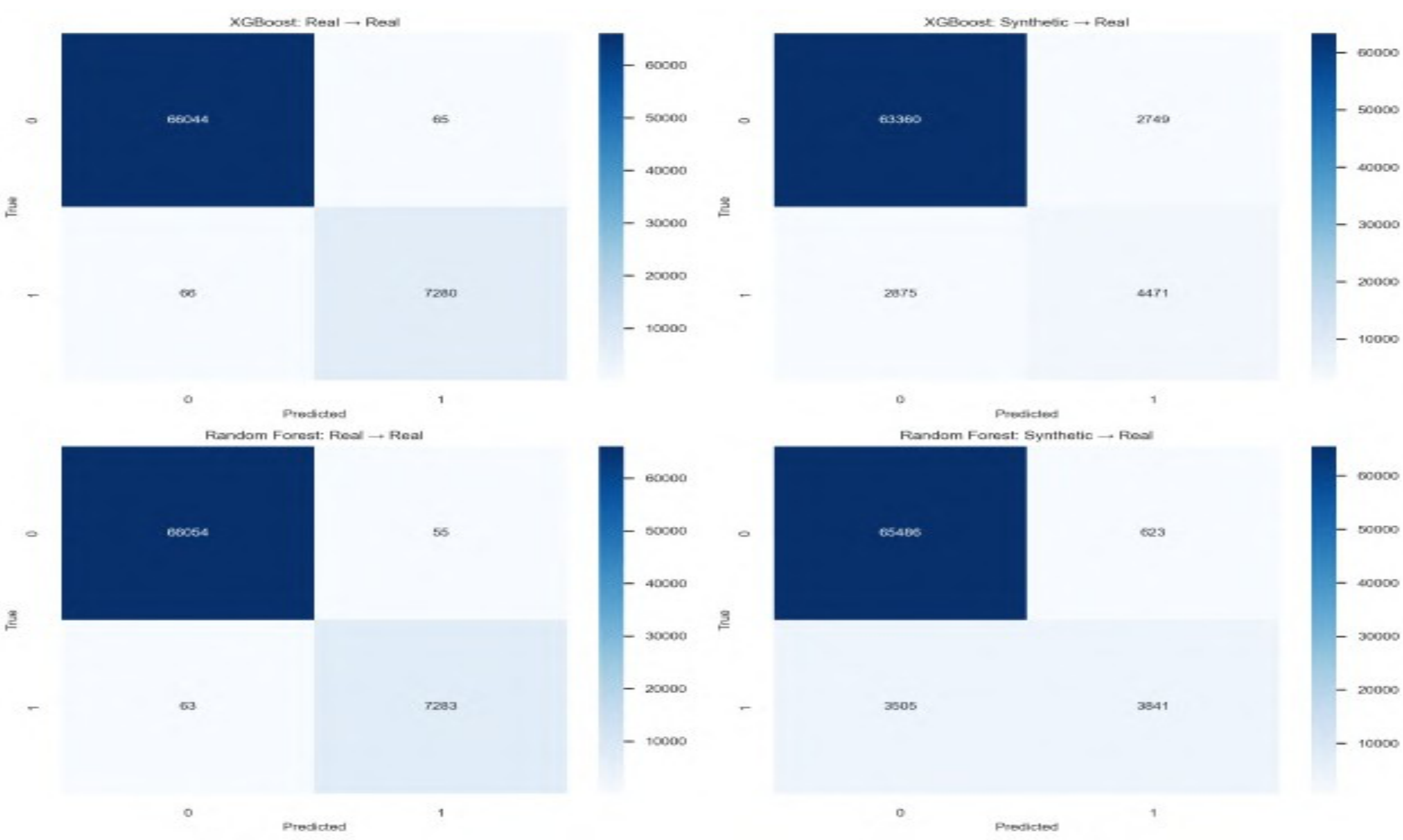

Fig. 11: Confusion matrices for different training scenarios. Top: XGBoost trained on real (left) vs. synthetic (right) data. Bottom: Random Forest trained on real (left) vs. synthetic (right) data.

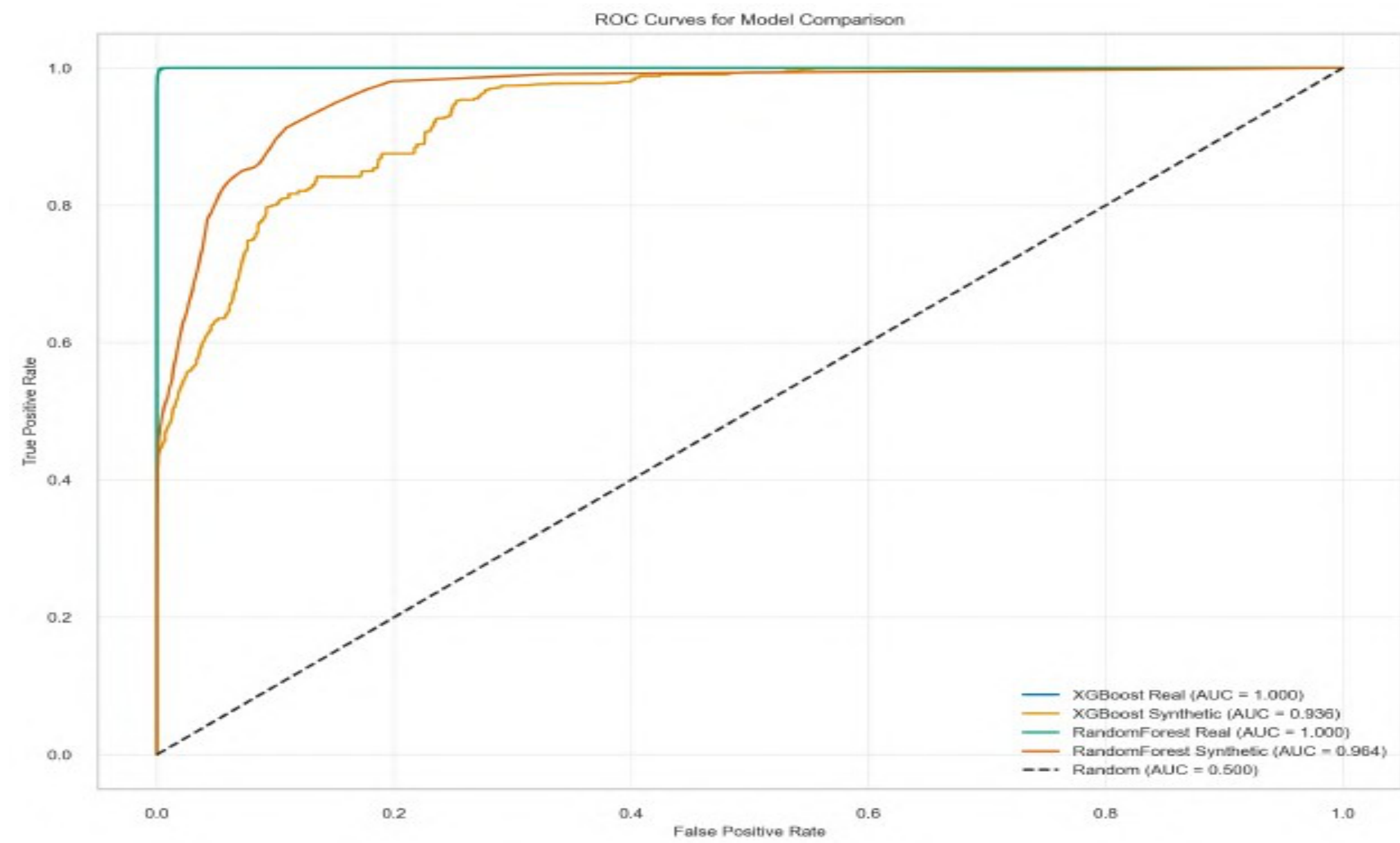

Fig. 12: ROC curves for different models and training scenarios. Models trained on synthetic data (dashed lines) show competitive AUC scores compared to those trained on real data (solid lines).

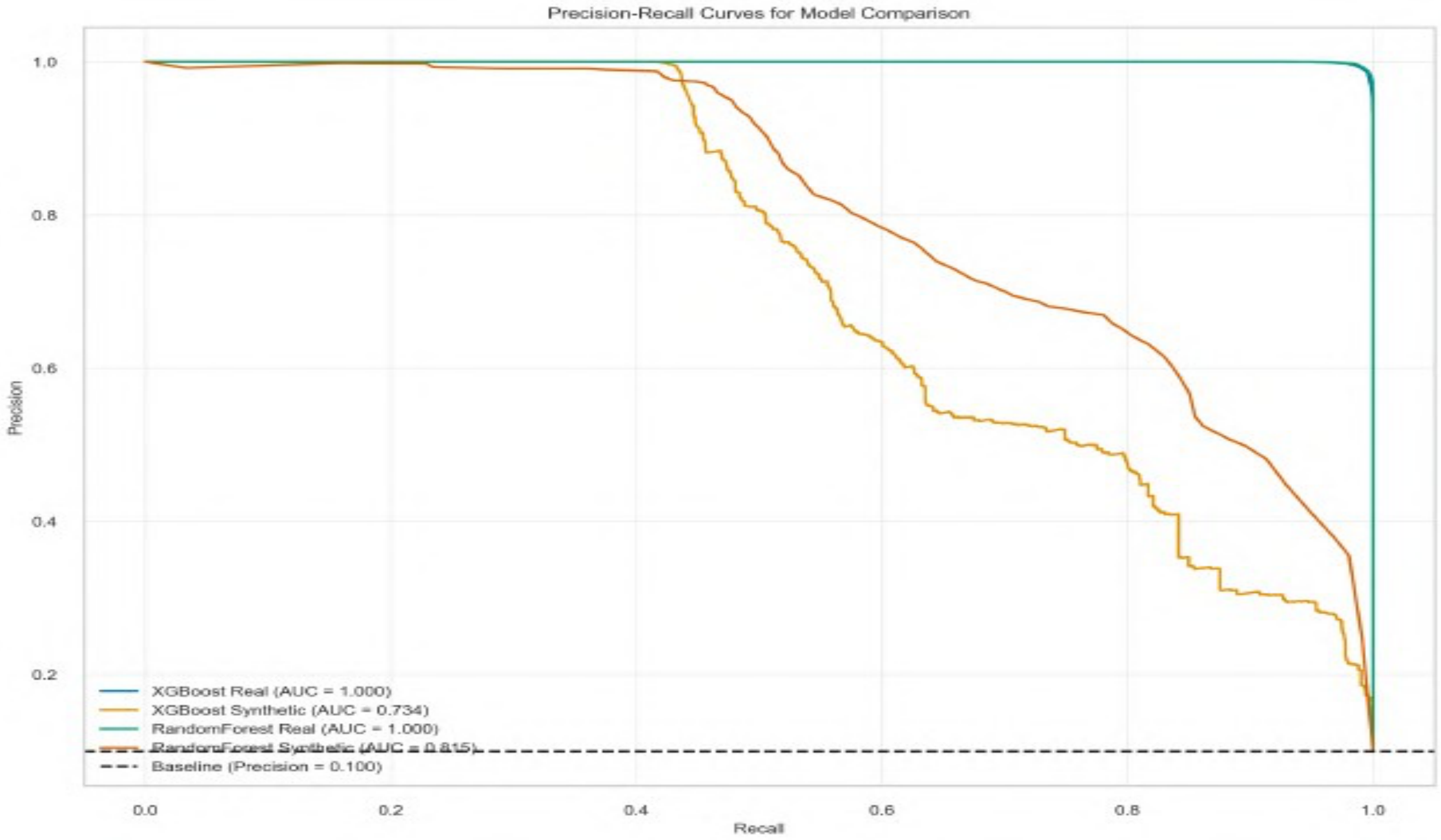

Fig. 13: Precision-Recall curves for different model and training scenarios. The curves illustrate the trade-off between precision and recall, with synthetic-trained models showing good performance at high recall operating points.

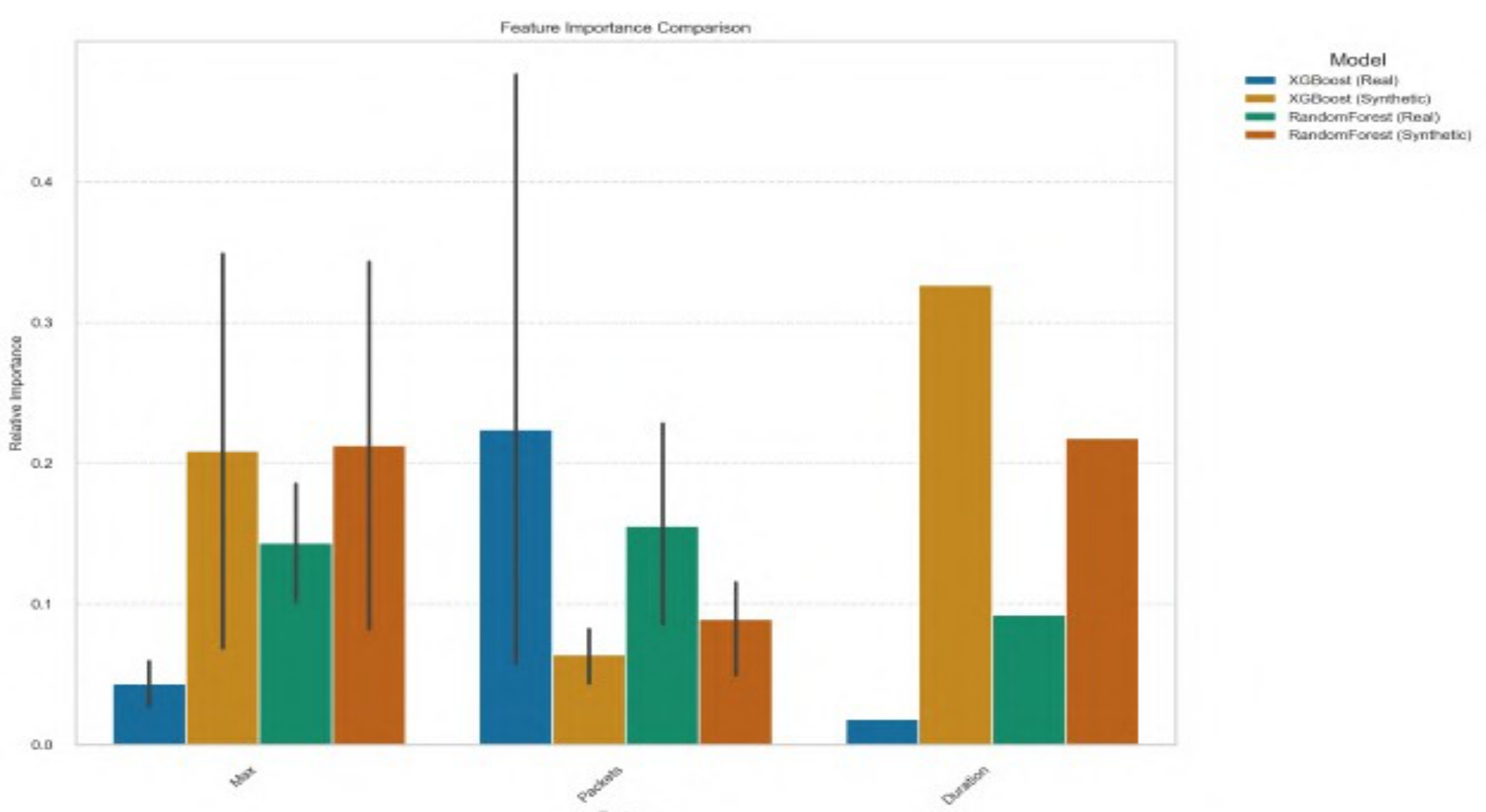

Fig. 14: Feature importance comparison across models and training scenarios. Similar importance patterns between real and synthetic-trained models indicate that synthetic data pre- serves the discriminative value of key features.

The ROC curves in Fig. 12 provide a more detailed view of model performance across different classification thresholds. The AUC scores for models trained on synthetic data remain competitive, particularly for tree-based models, further con- firming the utility of our synthetic data. The Precision-Recall curves in Fig. 13 reveal that models trained on synthetic data maintain good recall but experience some degradation in precision, particularly at high recall operating points. This suggests that while our synthetic data captures most anomaly patterns, there remain some subtle characteristics that are challenging to replicate. Analysis of feature importance (Fig. 14) shows consistent patterns between models trained on real and synthetic data, indicating that our synthetic data preserves the discriminative value of key features. The Flow Duration and packet length features emerge as particularly important across all models. These results suggest that while our synthetic data captures most of the patterns present in real traffic, there remain subtle characteristics of anomalous traffic that are challenging to reproduce synthetically. This finding aligns with previous research highlighting the complexity of generating realistic anomalous network traffic [7]. Our comprehensive evaluation demonstrates that the pro- posed synthetic data generation approach successfully pre- serves many critical aspects of encrypted network traffic data, including feature distributions and correlations. The relative performance of classifiers trained on synthetic data reached approximately 93% of the performance of those trained on real data, indicating that our synthetic data captures most of the discriminative patterns present in real network traffic.

Several key observations emerge from our results:

- **Preservation of Normal Traffic Patterns**: Our synthetic data generator was highly

effective at capturing and reproducing normal traffic patterns, as evidenced by the high accuracy for class 0 predictions.

- **Challenges with Anomaly Reproduction**: Generating synthetic anomalies that fully capture the complexity of real network attacks remains challenging. While our approach produced useful anomaly samples, there are still subtle characteristics of real anomalies that are difficult to replicate.
- **Model-Specific Sensitivity:** Different model architectures showed varying sensitivity to the use of synthetic training data. Tree-based models (XGBoost and Random Forest) demonstrated greater robustness to synthetic data than neural networks, suggesting that the choice of model architecture is important when working with synthetic data.
- **Correlation Preservation Success:** Our correlation- aware generation approach successfully maintained the relationship structure between features, as confirmed by the correlation matrix comparison.
- **Statistical Fidelity:** The statistical properties of the syn- thetic data closely matched those of the real data, with high similarity scores for means, standard deviations, skewness, and kurtosis as demonstrated in our quality assessment dashboard.

  The slightly lower performance of models trained on synthetic data compared to real data highlights the inherent tradeoff between data privacy (achieved through synthetic data) and model performance. However, the substantial performance achieved with synthetic data suggests that it can be a valuable complement to real data, especially in scenarios where real anomalous traffic samples are limited.

## VI. Conclusion

This paper presented a comprehensive approach to generate synthetic encrypted network traffic data using Generative AI techniques. Our methodology combined statistical analysis, clustering, and correlation-aware data generation to create synthetic datasets that preserve the key characteristics of real network traffic while enabling better balance between normal and anomalous traffic classes. The evaluation results demonstrate that our approach successfully captures most of the patterns present in real encrypted traffic, with classifiers trained on synthetic data achieving up to 93% of the performance of those trained on real data. This suggests that synthetic data can be a valuable complement to real data for training network security models, particularly in scenarios where anomalous traffic samples are limited. Key contributions of this work include:

- A comprehensive methodology for analyzing and preserving feature distributions and correlations in encrypted traffic data
- A cluster-based synthetic data generation approach that maintains the diversity of traffic patterns
- A rigorous evaluation framework combining statistical and predictive measures of synthetic data quality
- A quality assessment dashboard that provides an objective measure of synthetic data fidelity

Future work could explore more advanced generative models such as conditional GANs or VAEs for encrypted traffic generation, potentially improving the fidelity of generated anomalies. Additionally, investigating the transferability of synthetic data across different network environments could provide valuable insights into the generalizability of models trained on synthetic data.

DATA & CODE AVAILABILITY: https://github.com/harshilpatel22/Synthetic-Traffic-Generator